\documentclass[aps,prl,twocolumn, amssymb, amsmath, longbibliography, superscriptaddress]{revtex4-1}
\usepackage[latin9]{inputenc}
\usepackage{amsmath}
\usepackage{graphicx}

\makeatletter

\usepackage{dcolumn}
\usepackage{bm}
\usepackage{color}
\usepackage[makeroom]{cancel}

\makeatother

\begin{document}
\global\long\def\e{\varepsilon}%
\global\long\def\off{\text{off}}%
\global\long\def\fb{\text{fb}}%
\global\long\def\av{\text{av}}%
\global\long\def\l{\text{l}}%
\global\long\def\PC{\text{PC}}%

\title{Dynamical learning of dynamics}
\author{Christian Klos}
\affiliation{Neural Network Dynamics and Computation, Institute of Genetics, University
of Bonn, Bonn, Germany.}
\author{Yaroslav Felipe Kalle Kossio}
\affiliation{Neural Network Dynamics and Computation, Institute of Genetics, University
of Bonn, Bonn, Germany.}
\author{Sven Goedeke}
\affiliation{Neural Network Dynamics and Computation, Institute of Genetics, University
of Bonn, Bonn, Germany.}
\author{Aditya Gilra}
\affiliation{Neural Network Dynamics and Computation, Institute of Genetics, University
of Bonn, Bonn, Germany.}
\affiliation{Department of Computer Science, and Neuroscience Institute, University
of Sheffield, Sheffield, UK.}
\author{Raoul-Martin Memmesheimer}
\affiliation{Neural Network Dynamics and Computation, Institute of Genetics, University
of Bonn, Bonn, Germany.}
\begin{abstract}
The ability of humans and animals to quickly adapt to novel tasks
is difficult to reconcile with the standard paradigm of learning by
slow synaptic weight modification. Here we show that fixed-weight
neural networks can learn to generate required dynamics by imitation.
After appropriate weight pretraining, the networks quickly and dynamically
adapt to learn new tasks and thereafter continue to achieve them without
further teacher feedback. We explain this ability and illustrate it
with a variety of target dynamics, ranging from oscillatory trajectories
to driven and chaotic dynamical systems.
\end{abstract}
\maketitle
\textit{Introduction.} Humans and animals can learn wide varieties
of tasks. The predominant paradigm assumes that their neural networks
achieve this by slow adaptation of connection weights between neurons
\citep{DA01,gerstner2014neuronal}. Neurobiological experiments, however,
also indicate fast learning with static weights \citep{perich2018neural}.
Our study addresses how neural networks may quickly learn to generate
required output dynamics without weight-learning.

The goal of neural network learning is ultimately to appropriately
change the activity of the output neurons of the network. In supervised
learning, it should match a target and continue doing so during subsequent
testing; in reinforcement learning, it should maximize a sparsely
given reward. In our study, the networks adapt their weights during
a pretraining phase \citep{thrunLearningLearn1998,vanschorenMetaLearningSurvey2018,lansdellLearningtolearn2018}
such that thereafter with static weights they achieve supervised learning
of desired outputs, by adapting only their dynamics (dynamical learning).
Adapting the static network's weights during pretraining is thus a
kind of meta-learning or learning-to-learn. There is a recent spurt
of interest in learning-to-learn \citep{vanschorenMetaLearningSurvey2018,lansdellLearningtolearn2018},
focusing mainly on learning of reinforcement learning \citep{duanRLFastReinforcement2016,wangPrefrontalCortexMetareinforcement2018,nagabandiLearningAdaptMetaLearning2018}.
Studies on learning of supervised dynamical learning showed prediction
of a time series at the current time step given the preceding step's
target \citep{feldkampAdaptationFixedWeight1996,feldkampAdaptiveBehaviorFixed1997,youngerFixedweightOnlineLearning1999,hochreiterLearningLearnUsing2001a,youngerMetalearningBackpropagation2001,feldkampSimpleConditionedAdaptive2003,santiagoContextDiscerningMultifunction2004,lukoseviciusEchoStateNetworks2007,santoroMetaLearningMemoryAugmentedNeural2016,bellecLongShorttermMemory2018}
and control of a system along a time-varying target \citep{feldkampTrainingRobustNeural1994,feldkampTrainingControllersRobustness1994,feldkampFixedweightControllerMultiple1997,oubbatiNeuralFrameworkAdaptive2010}.
The studies assume that a target is present during testing to avoid
unlearning. This limits applicability and renders the dynamics necessarily
non-autonomous; it is conceptually problematic for supervised settings
and at odds with the common concept of teacher-free recall.

We therefore develop a scheme for fast supervised dynamical learning
and subsequent teacher-free generation of long-term dynamics. We consider
models for biological recurrent (reciprocally connected) neural networks,
where leaky rate neurons interact in continuous time \citep{DA01,gerstner2014neuronal}.
Such models are amenable to learning, computation and phase space
analysis \citep{DA01,gerstner2014neuronal,JLP+07,SA09,sussillo2013opening}.
After appropriate pretraining using the reservoir computing scheme
(where only the weights to output neurons are trained \citep{Supplement19DynLearn,JH04,MNM02,SA09}),
all weights are fixed. The networks can nevertheless learn to generate
new, desired dynamics. Furthermore, they continue to generate them
in self-contained manner during subsequent testing. We illustrate
this with a variety of trajectories and dynamical systems and analyze
the underlying mechanisms.

\emph{Network model. }We use recurrent neural networks, where each
neuron (or neuronal subpopulation) $i,\,i=1,...,N$, with $N$ between
500 and 3000 depending on the task \citep{Supplement19DynLearn},
is characterized by an activation variable $x_{i}(t)$ and communicates
with other neurons via its firing rate $r_{i}(t)$, a nonlinear function
of $x_{i}(t)$ \citep{DA01,gerstner2014neuronal}. In isolation $x_{i}(t)$
decays to zero with a time constant $\tau_{i}$. This combines the
decay times of membrane potential and synaptic currents. The network
has two outputs, which can be interpreted as linear neurons:
\begin{figure}
\noindent \begin{centering}
\includegraphics[width=1\columnwidth]{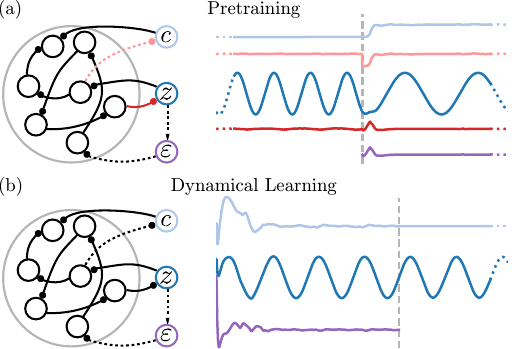}
\par\end{centering}
\caption{\label{fig:network structure}Pretraining and learning. (a) During
pretraining, the output weights (left, different reds) of the network
are adapted using the output errors $\protect\e(t)=z(t)-\tilde{z}(t)$
(right, red) and $c(t)-\tilde{c}(t)$ (light red), such that $z(t)$
(blue, different scale for clarity) and $c(t)$ (light blue) match
their targets. Different members of the target family are weight-learned
in the training periods (dashed vertical). At their beginnings, $\protect\e(t)$
is fed also as input (purple). (b) Dynamical learning. The output
weights are now fixed. The network receives the signal error $\protect\e(t)$
as input (purple). It adapts its dynamics to generate $z(t)\approx\tilde{z}(t)$
(blue). During testing, an error is no longer provided and $c(t)$
is fixed to its previous average (right, dashed vertical, left, dashed
weights). $z(t)$ continues to approximate $\tilde{z}(t)$.}
\end{figure}
 signal $z_{k}(t),\,k=1,...,N_{z}$, and context $c_{l}(t),\,l=1,...,N_{c}$
(Fig.~\ref{fig:network structure}). After learning, $z(t)$ generates
the desired dynamics while $c(t)$ indexes it. They are continually
fed back to the network, allowing their autonomous generation \citep{JH04}.
The networks are temporarily also informed about their signal's difference
from its target $\tilde{z}(t)$ by an error input $\e(t)=z(t)-\tilde{z}(t)$.
Taken together, for constant weights the network dynamics are given
by
\begin{equation}
\begin{array}{rl}
\tau\dot{x}(t) & \!\!=-x(t)+Ar(t)+w_{z}z(t)+w_{c}c(t)\\
 & \!\phantom{=}+w_{\e}\e(t)+w_{u}u(t),\\
z(t) & \!\!=o_{z}r(t),\,c(t)=\text{\ensuremath{o_{c}}}r(t),
\end{array}\label{eq:x dyn}
\end{equation}
with recurrent weights $A$, the diagonal matrix of time constants
$\tau$, signal and context output weights $o_{z},\,o_{c}$, feedback
weights $w_{z},\,w_{c}$, input weights $w_{\e},\,w_{u}$ and a drive
$u(t)$ absent for most tasks. We choose $r_{i}(t)=\tanh(x_{i}(t)+b_{i})$
\citep{JH04,SA09,lukosevicius2012reservoir}; offsets $b_{i}$ are
drawn from a uniform distribution between $-0.2$ and $0.2$ and break
the $x\rightarrow-x$ symmetry without input. Unless mentioned otherwise,
we set $\tau_{i}=1$ fixing the overall time scale. Recurrent weights
$A_{ij}$ are set to zero with probability $1-p$ ($p=0.1\text{ or }p=0.2$
depending on the task). Nonzero weights are drawn from a Gaussian
distribution with mean $0$ and variance $\frac{g^{2}}{pN}$, where
$g=1.5$ \citep{SA09}. $w_{z,ij},$ $w_{c,ij}$ and $w_{\varepsilon,ij},$
$w_{u,ij}$ are drawn from a uniform distribution between $-\tilde{w}$
and $\tilde{w}$ ($\tilde{w}=1\text{ or }\tilde{w}=2$).

\emph{Pretraining.} The aim of our pretraining (Fig.~\ref{fig:network structure}a)
is twofold. First, it should enable the resulting static networks
to learn signals of a specific class given only the error input $\e(t)$.
Second, after removing the error input the static networks should
be able to continue to generate the desired dynamics. Therefore, the
networks have to learn to minimize $\e(t)$ and, as explained in the
Analysis section, to associate unique contexts with the different
target dynamics.

To achieve this, we present different trajectories $\tilde{z}(t)$
of the target class to the networks, together with associated, straightforwardly
chosen constant indices $\tilde{c}$. The output weights $o_{z,ij}$
and $o_{c,ij}$ learn online according to the FORCE rule \citep{Supplement19DynLearn,SA09}
to minimize the output errors $\e(t)$ and $c(t)-\tilde{c}$. In short,
they are modified using the supervised recursive least-squares algorithm
with high learning rate. This provides a least-squares optimal regularized
solution for the output weights given the past network states and
targets \citep{ismail1996equivalence}. Signals and indices are presented
for a time $t_{\text{wlearn}}$ (30000 or 50000) as a continuous,
randomly repeating sequence of training periods of duration $t_{\text{stay}}$
(between 200 and 1000). During each training period's first part,
a network receives $\e(t)$ as input. Because of the various last
states of the previous learning periods, it thus learns to approach
$\tilde{z}(t)$ from a broad range of initial conditions given this
input. In most of our tasks, after a time $t_{\fb}=100$, when $z(t)$
is close to $\tilde{z}(t)$, $\e(t)$ is switched off and $c(t)$
is fixed to its constant target, matching the testing paradigm. This
often helps the network to learn generation of $z(t)\approx\tilde{z}(t)$
without error input.

\emph{Dynamical learning and testing.} The weights now remain static
and the error input teaches the network new tasks of the pretrained
target class (Fig.~\ref{fig:network structure}b), i.e.~the networks
dynamically learn to generate $z(t)\approx\tilde{z}(t)$ for previously
unseen $\tilde{z}(t)$. The learning time $t_{\text{learn}}$ (between
50 and 200) is short, a few characteristic timescales of the target
dynamics \citep{Supplement19DynLearn}. $c(t)$ is moderately fluctuating.

Thereafter the test phase begins, where no more teacher is present
($w_{\e}\rightarrow0$). In weight-learning paradigms, during such
phases the weights are fixed \citep{MNM02,JH04,SA09,mante2013context,hennequin2014optimal}.
We likewise fix $c(t)$ to a temporally constant value, an average
of previously assumed ones, $c(t)=\bar{c}$. This may be interpreted
as indicating that the context is unchanged and the same signal is
still desired. We find in our applications, that the network dynamics
continue to generate a close-to-desired signal $z(t)$, establishing
the successful dynamical learning of the task.

\emph{Applications.} We illustrate our approach by learning a variety
of trajectories (tasks (i-iv)) and dynamical systems (tasks (v,vi)).
First, we consider a family $\tilde{z}(t;k)$ of target trajectories,
parameterized by $k$. The networks are pretrained on a few of them,
where the context target $\tilde{c}$ is a linear function of $k$.
Thereafter the networks dynamically learn to generate a previously
unseen trajectory as output and perpetuate it during testing. We start
with the simple, instructive target family of oscillations with different
periods (task (i)): $\tilde{z}(t;T)=5\sin(\frac{2\pi}{T}t)$. We use
three different teacher trajectories for pretraining, with $T=10,15,20$.
After pretraining, our networks can precisely dynamically learn oscillations
with unseen periods within and slightly beyond the pretrained ones
(Fig.~\ref{fig:applications}a,b, see \citep{Supplement19DynLearn}
for further detail and analysis of learning performance of all tasks).
Next, in (ii), we generalize (i) to higher order Fourier series. Specifically,
we consider the target family of superpositions of two random Fourier
series with weighting factor $\lambda$: $\tilde{z}(t;\lambda)=(1-\lambda)\tilde{z}_{1}(t;\lambda)+\lambda\tilde{z}_{2}(t;\lambda)$.
Here, $\tilde{z}_{l}(t;\lambda),\,l=1,2,$ are Fourier series of order
$O$ and period $T(\lambda)=(1-\lambda)T_{1}+\lambda T_{2}$. $T_{l}$
and the Fourier coefficients are drawn randomly. We use seven different
teacher trajectories for pretraining, with weighting factors distributed
equidistantly between 0 and 1. After pretraining, we test the dynamical
learning for thirteen weighting factors also distributed equidistantly
between 0 and 1. To quantify the learning performance, we determine
the fraction of these targets that can be successfully learned (RMSE
below given threshold (0.4) and below RMSE between signal and (other)
pretrained targets). We find that networks of increasing size can
learn Fourier series with increasing order (Fig.~\ref{fig:applications}c,d).
Networks with 3000 neurons learn Fourier series of order 10 with a
median fraction of successes of close to $90\%$. Hence, very general
periodic functions can be learned. The highest producible frequency
is limited by the available neuronal time scales $\tau_{i}$. We thus
expect that larger networks containing smaller $\tau_{i}$ can learn
even higher order targets.

\begin{figure}
\includegraphics{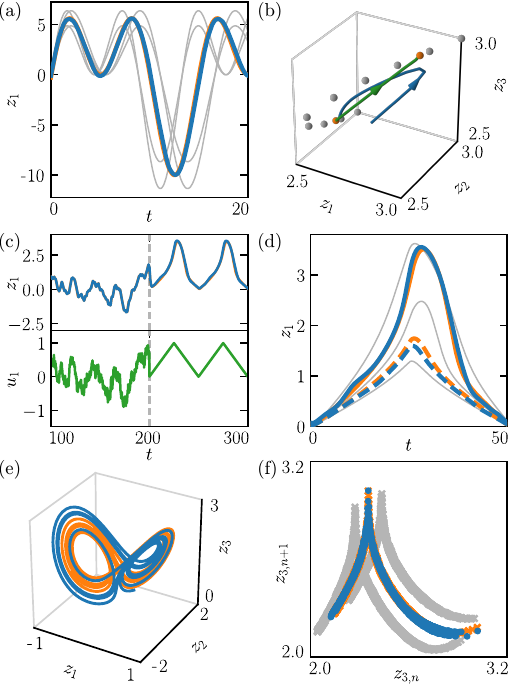}

\caption{\label{fig:further_applications}Dynamical learning of different tasks.
Testing phase after dynamical learning of an example (a) two-parametric
superposition of sines, (b) fixed point, (c,d) driven overdamped pendulum,
and (e,f) Lorenz system. (a-d) Signals (blue) match testing targets
(orange, mostly covered by signal) well. Pretraining targets (gray
traces or spheres) are clearly distinct. (a) displays only the four
closest pretrained dynamics for clarity. (b) Signal transients (blue,
green) of subsequent dynamical learning of two targets (orange spheres).
(c) Signal, target and drive (green) during dynamical learning and
subsequent testing (dashed vertical). (d) Dynamically learned approximations
of two different pendulums (continuous, dashed), driven by the same
triangular input. (e) Limit sets of signal (blue) and target (orange).
(f) Tent maps of signal (blue) and dynamical (orange) and pretrained
targets (gray).}
\end{figure}

To check if our approach also works for a target family with more
than one parameter and multidimensional trajectories, we consider
in (iii) a superposition of sines with different amplitude and period
(consequently $k,\tilde{c}$ are two-dimensional vectors) and in (iv)
a set of fixed points along a curve in three-dimensional space. We
find that, after pretraining, our networks are able to dynamically
learn unseen members of these target families with multidimensional
context or signal, as shown in Fig.~\ref{fig:further_applications}a,b
for example trajectories. 

Second, we consider a family $\dot{\tilde{z}}(t)=F(\tilde{z}(t),u(t);k)$
of target dynamical systems. The networks are pretrained on a few
representative systems. Thereafter, an unseen one is dynamically learned.
Learning is in both phases based on imitation of trajectories. However,
in contrast to tasks (i-iv) the networks now need to generate unseen
output trajectories during testing. To demonstrate dynamical learning
of a driven system, we consider task (v) of approximating the trajectory
of an overdamped pendulum with drive $u(t)$ and different masses
$m$: $\dot{\tilde{z}}(t)=F(\tilde{z}(t),u(t);m)$. During pretraining
and dynamical learning, we use low-pass filtered white noise as drive
(Fig.~\ref{fig:further_applications}c, left of dashed vertical).
During testing, we use a triangular wave (Fig.~\ref{fig:further_applications}c,
right of dashed vertical). As our networks nevertheless generate the
correct qualitatively different signal (Fig.~\ref{fig:further_applications}c,d),
they must have learned the underlying vector field $F(\tilde{z},u;m)$.
(v) also shows that learning goes beyond interpolation of trajectories
(compare blue and gray traces in Fig.~\ref{fig:further_applications}d).
Finally, in task (vi) we show dynamical learning of chaotic dynamics,
considering autonomous Lorenz systems with different dissipation parameter
$\beta$ of the $z$-variable. For chaotic dynamics, even trajectories
of similar systems quickly diverge. The aim in this task is thus only
to generate during testing signals of the same type as the trajectories
of the target system. We test this by comparing the limit sets of
the dynamics and the tent-map relation between subsequent maxima of
the $z$-coordinate (Fig.~\ref{fig:further_applications}e,f). The
reproduction of the tent-map relation further shows that our approach
can generate not explicitly trained quantitative dynamical features.
We note that the networks also dynamically learn the fixed point convergence
of some of the targets in the considered parameter space, even though
they were pretrained on chaotic dynamics only \citep{Supplement19DynLearn}.

\begin{figure}
\includegraphics{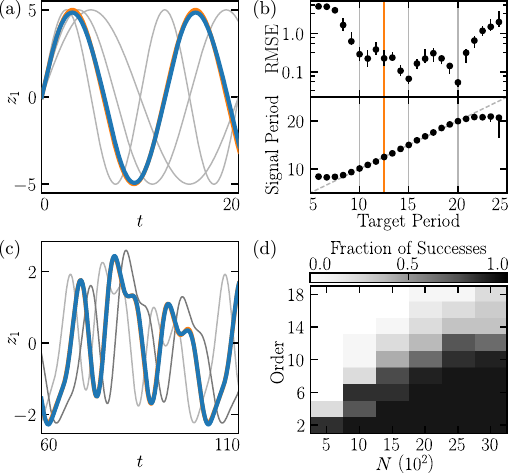}

\caption{\label{fig:applications}Dynamical learning of periodic trajectories.
(a,b) Testing after dynamical learning of sinusoids. (a) Signal (blue)
matches the example testing target (orange, mostly covered by signal)
well. Pretraining targets (gray traces) are clearly distinct. (b)
For many different targets the root-mean-square error (RMSE) between
signal and target is low (top) and the signal's period tracks the
target's period well (bottom). Gray and orange verticals indicate
periods of pretrained targets and target in (a). Dots show median
value and errorbars interquartile range, using 10 network instances.
(c,d) Testing after dynamical learning of Fourier series. (c) Like
(a), for a random Fourier series with $O=6$. Only the two closest
pretrained dynamics are displayed for clarity. (d) Learning success
for different network sizes and orders of the Fourier series. Color
encodes median fraction of success, using 40 network instances and
random Fourier series.}
\end{figure}

\emph{Analysis. }In the following we analyze the different parts of
our network learning and its applicability. One interpretation of
the pretraining phase is that the network learns a negative feedback
loop, which reduces the error $\e(t)$. For another interpretation,
we split $\e(t)$ and regroup the $z$-dependent part of Eq.~\eqref{eq:x dyn}
as $(w_{z}+w_{\e})z(t)-w_{\e}\tilde{z}(t)$: feeding back $\e(t)$
is equivalent to adding a teacher drive $\tilde{z}(t)$, except for
a specific change in the feedback weights $w_{z}$. For the $z$-output
alone the network thus weight-learns an autoencoder $\tilde{z}(t)\rightarrow z(t)$.
This is usually an easy task for reservoir networks \citep{abbott2016building}.
To simultaneously learn the constant output $c(t)=\tilde{c}$, the
network has to choose an appropriate $o_{c}$ orthogonal to the subspaces
in which the different $z(t)$-driving $r$-dynamics take place. Orthogonal
directions are available in sufficiently large networks, since the
subspaces are low-dimensional \citep{abbott2011interactions}.

After the correct $z$-dynamics are assumed, we have $\e(t)\approx0$.
Since remaining fluctuations in $\e(t)$ could stabilize the dynamics,
we usually include ensuing learning phases with $w_{\e}\rightarrow0$
and $c(t)=\tilde{c}$. These teach the network to generate the correct
dynamics in stable manner under conditions similar to testing.

\begin{figure}
\noindent \begin{centering}
\includegraphics{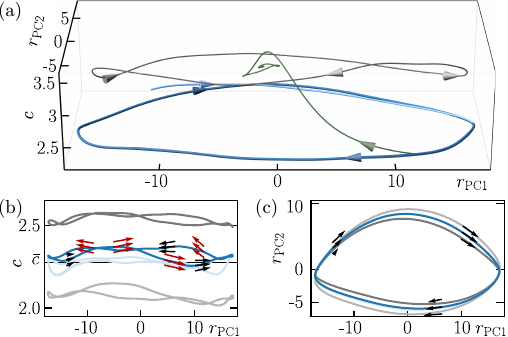}
\par\end{centering}
\caption{\label{fig:analysis figure} Network dynamics during dynamical learning
(a) and testing (b,c) of task (i), in $c,r_{\protect\PC1},r_{\protect\PC2}$-coordinates.
(a) During dynamical learning, the error input drives the network
to a periodic orbit (light blue trajectory) and keeps it there (blue).
Without input, the dynamics converge to a stable orbit (gray) whose
signal approximates a pretrained one. Freezing $\tilde{z}(t)=\tilde{z}(t_{0})$
drives the dynamics to a fixed point off the orbit (green). (b) During
testing, the assumed orbit (blue) in the $c\text{-}r_{\protect\PC,1}$-plane
is similar to the error driven one (light blue, closest pretrained
orbits with $c(t)$ fixed to their $\tilde{c}$: gray). The constant
feedback $\bar{c}$ prevents the dynamics to leave the region where
$c(t)\approx\bar{c}$, compare $\dot{r}(r)$ (black vectors, $r$
on/nearby trajectory) with $\dot{r}(r)$ for variable feedback $c(t)$
(red vectors). (c) All four orbits are similar in the $r_{\protect\PC,1}\text{-}r_{\protect\PC,2}$-plane.
The dynamically learned orbit has an attracting projected vector field
(black vectors) like the pretrained orbits.}
\end{figure}

To analyze the principles underlying dynamical learning and testing,
we consider task (i). The similarity of the network and learning setups
suggests that the same principles underlie all our tasks. We additionally
confirm this for (vi) \citep{Supplement19DynLearn}. Viewing the network
dynamics in the space of firing rates $r$, we choose new coordinates
with first axis along $o_{c}$ and the principal components of the
dynamics orthogonal to $o_{c}$. The dynamics are then given by $c(t)=o_{c}r(t),r_{\PC1}(t),r_{\PC2}(t),...$
(Fig.~\ref{fig:analysis figure}). We focus on the first three coordinates,
which describe large parts of the dynamics and output generation.
 We find that during dynamical learning, the error feedback drives
the dynamics towards an orbit that is shifted in $c$ but similar
to pretrained ones. The network therewith generalizes the pretrained
reaching and generation of orbits together with corresponding, near-constant
$c(t)$, while $\e(t)$ is fed in. We note that the combination of
current state and error input is important (see Fig.~\ref{fig:analysis figure}a
for $w_{\e}\rightarrow0$ and a mismatched $\tilde{z}(t)=\tilde{z}(t_{0})\,\,\text{for}\,\,t>t_{0}$).

During testing, the network generalizes the pretrained characteristics
that feeding back $w_{c}\tilde{c}$ leads to $c(t)\approx\tilde{c}$.
Clamping $w_{c}c(t)\,\,\text{to}\,\,w_{c}\bar{c}$ thus results in
an approximate restriction of $r(t)$ to an $N-1$-dimensional hyperplane
with $c(t)=o_{c}r(t)\approx\bar{c}$ (Fig.~\ref{fig:analysis figure}b).
The resulting trajectory is for task (i) a stable periodic orbit that
generates the desired signal, because the vector field projected to
the $c(t)=\bar{c}$-hyperplane is similar to the vector field projected
to the $c(t)=\tilde{c}$-hyperplanes embedding nearby pretrained periodic
orbits (Fig.~\ref{fig:analysis figure}c).

\emph{Discussion and conclusion. }We have introduced a scheme how
neural networks can quickly learn dynamics without changing their
weights and without requiring a teacher during testing. It relies
on a weight-learned mutual association, quasi an entanglement, between
contexts and targets. This enables the latter to fix the former during
dynamical learning and vice versa during testing.

Previous approaches to supervised dynamical learning with continuous
signal space required a form of the teaching signal also during testing.
They further differ in network architecture, learning algorithm, task
and/or assumption of discrete time from ours \citep{Supplement19DynLearn,feldkampTrainingRobustNeural1994,feldkampTrainingControllersRobustness1994,feldkampAdaptationFixedWeight1996,feldkampFixedweightControllerMultiple1997,feldkampAdaptiveBehaviorFixed1997,youngerFixedweightOnlineLearning1999,hochreiterLearningLearnUsing2001a,youngerMetalearningBackpropagation2001,feldkampSimpleConditionedAdaptive2003,santiagoContextDiscerningMultifunction2004,lukoseviciusEchoStateNetworks2007,oubbatiNeuralFrameworkAdaptive2010,santoroMetaLearningMemoryAugmentedNeural2016,bellecLongShorttermMemory2018,oubbatiMetaLearningAdaptiveIdentification2005,jaeger2008cant,wyffels2014frequency}.
In networks with external input unseen, interpolating input can lead
to interpolating dynamics \citep{Supplement19DynLearn,bostrom2013model}.
In contrast, our networks \emph{learn} new dynamics, by imitation.

Our scheme is conceptually independent of the network and weight-learning
model. The pretraining implements a form of structure learning, i.e.~learning
of the structure underlying a task family \citep{lansdellLearningtolearn2018,Braun2010}.
Animals and humans employ it frequently, but little is known about
its neurobiology. We thus realize it by a simple reservoir computing
scheme with FORCE learning \citep{Supplement19DynLearn,SA09}. We
checked that we can use biologically more plausible weight perturbation
learning for a simple fixed point learning task \citep{Supplement19DynLearn}.

Dynamical learning is biologically plausible: it is naturally local,
causal and does not require fast synaptic weight updates. Continuous
supervision could be generated by an inverse model \citep{jordan1992forward}
and might be replaceable by a sparse, partial signal. Our dynamical
learning is fast \citep{Supplement19DynLearn} (cf\@.~also \citep{wangPrefrontalCortexMetareinforcement2018,jaeger2008cant,feldkampAdaptationFixedWeight1996,feldkampAdaptiveBehaviorFixed1997,hochreiterLearningLearnUsing2001a,youngerMetalearningBackpropagation2001}).
Even for more complicated tasks convergence requires only a few multiples
of a characteristic time scale of the dynamics. Further, we find robustness
against changes in network and task parameters \citep{Supplement19DynLearn}.
The above points suggest a high potential of our scheme for applications
in biology, physics and engineering such as neuromorphic computing
and the prediction of chaotic systems \citep{Supplement19DynLearn}.
\begin{acknowledgments}
We thank Paul Züge for fruitful discussions, the German Federal Ministry
of Education and Research (BMBF) for support via the Bernstein Network
(Bernstein Award 2014, 01GQ1710) and the European Union's Horizon
2020 research and innovation programme for support via the Marie Sk\l odowska-Curie
Grant No 754411.
\end{acknowledgments}

%

\end{document}


\global\long\def\e{\varepsilon}%
\global\long\def\off{\text{off}}%
\global\long\def\fb{\text{fb}}%
\global\long\def\av{\text{av}}%
\global\long\def\l{\text{l}}%
\global\long\def\PC{\text{PC}}%

\title{Dynamical learning of dynamics\\
\textendash{} Supplemental Material \textendash{}}
\author{Christian Klos}
\affiliation{Neural Network Dynamics and Computation, Institute of Genetics, University
of Bonn, Bonn, Germany.}
\author{Yaroslav Felipe Kalle Kossio}
\affiliation{Neural Network Dynamics and Computation, Institute of Genetics, University
of Bonn, Bonn, Germany.}
\author{Sven Goedeke}
\affiliation{Neural Network Dynamics and Computation, Institute of Genetics, University
of Bonn, Bonn, Germany.}
\author{Aditya Gilra}
\affiliation{Neural Network Dynamics and Computation, Institute of Genetics, University
of Bonn, Bonn, Germany.}
\affiliation{Department of Computer Science, and Neuroscience Institute, University
of Sheffield, Sheffield, UK.}
\author{Raoul-Martin Memmesheimer}
\affiliation{Neural Network Dynamics and Computation, Institute of Genetics, University
of Bonn, Bonn, Germany.}
\maketitle

\section{Reservoir computing and FORCE learning\label{sec:Reservoir-computing-and}}

Reservoir computing has been introduced several times at different
levels of elaborateness and in different flavors, in machine learning
and in neuroscience \citep{BM95,dominey1995complex,JH04,MNM02}. A
reservoir computer consists of a high-dimensional, nonlinear dynamical
system, the reservoir or liquid, and a comparably simple readout.
The reservoir, often a recurrent neural network, ``echoes'' the
input in a complicated, nonlinear way; it acts like a random filter
bank with finite memory as each of its units generates a nonlinearly
filtered version of the current input and its recent past while forgetting
more remote inputs \citep{BM95,westover2002linearly,MNM02,JH04}.
The simple, often linear readout can then be weight-trained to extract
the desired results while the reservoir remains static. Only a fraction
of the neural network weights are therefore used for task-related
adaptation.

We use a reservoir computing scheme for pretraining, see main text.
The output weights of our networks, the weights $o_{z}$ and $o_{c}$
to $z$ and $c$, learn online according to the FORCE rule \citep{SA09},
which is well suited for reservoir computers with output feedback
\citep{lukosevicius2012reservoir}. This is because it assumes fast
learning of the output weights with a powerful algorithm and thereby
ensures that the output and thus the feedback input always match the
desired ones up to a small error. The recurrent network is thus largely
driven by the correct feedback signals and generates appropriate dynamics
already during training. The remaining fluctuations are intrinsically
generated and therefore efficiently immunize the system against fluctuations
that will occur during testing, leading to dynamically stable generation
of desired dynamics. The output weights are trained using the supervised
recursive least-squares algorithm. This higher order algorithm provides
a least-squares optimal, usually regularized result given the past
network states and the targets. Concretely, the version used in ref.~\citep{SA09}
and in our article starts the recursion with $o_{z,ij}(0)$ and $o_{c,ij}(0)$
for the signal and context output weights and with an $N\times N$
matrix $P(0)=\alpha^{-1}I$, where $I$ is the identity matrix and
$\alpha^{-1}$ acts as a learning rate parameter. In learning step
$n$ at time $t_{n}$ the output weights $o_{z}(n)$ and $o_{c}(n)$
and the matrix $P(n)$ are recursively updated via
\begin{align}
o_{z,ij}(n) & =o_{z,ij}(n-1)-g_{j}(n)\varepsilon_{i}(t_{n}),\\
o_{c,ij}(n) & =o_{c,ij}(n-1)-g_{j}(n)e_{i}(t_{n}),\\
P(n) & =(I-g(n)r^{T}(t_{n}))P(n-1),
\end{align}
where $T$ denotes transposition, $r(t_{n})$ the outputs of the neurons
at time $t_{n}$ and $\varepsilon(t)=z(t)-\tilde{z}(t)$ and $e(t)=c(t)-\tilde{c}(t)$
the errors of the signal and the context. $g(n)=\left(1+r^{T}(t_{n})P(n-1)r(t_{n})\right)^{-1}P(n-1)r(t_{n})$
specifies the learning rates of $o_{z,ij}$ and $o_{c,ij}$. They
depend on the presynaptic neuron $j$ and on the dynamical history
of the entire reservoir, which renders the algorithm causal but non-local.
The recursion ensures that $o_{z}(n)$ and $o_{c}(n)$ minimize the
``ridge regression'' error functions
\begin{align}
E_{z,i}(n) & =\sum_{k=1}^{n}\left(\sum_{j}o_{z,ij}(n)r_{j}(t_{k})-\tilde{z}_{i}(t_{k})\right)^{2}+\alpha\sum_{j=1}^{N}\left(o_{z,ij}(n)-o_{z,ij}(0)\right)^{2},\\
E_{c,i}(n) & =\sum_{k=1}^{n}\left(\sum_{j}o_{c,ij}(n)r_{j}(t_{k})-\tilde{c}_{i}(t_{k})\right)^{2}+\alpha\sum_{j=1}^{N}\left(o_{c,ij}(n)-o_{c,ij}(0)\right)^{2},
\end{align}
i.e.~the individual signal and context errors are kept small with
weights that ideally do not deviate far from the initial ones (weight
regularization) \citep{ismail1996equivalence}. The non-locality and
the assumed fast weight changes are considered biologically implausible
\citep{SA09,Miconi2017}.

\clearpage{}

\section{Additional detail on the applications\label{sec:Additional-detail}}

In the following, we detail the parameters, setups and targets used
in the different applications. We denote the duration of the pretraining
phase by $t_{\text{wlearn}}$. Each training period (individual target
presentation) in it lasts for $t_{\text{stay}}$. If not mentioned
otherwise, in the beginning of each period until $t_{\text{fb}}$
the network receives error input $\varepsilon(t)$ and the context
signal evolves freely. Thereafter, $w_{\e}\rightarrow0$ and $c(t)$
is fixed to its target value. The intervals between updates of the
output weights have length $dt$ for task (ii) and random lengths
with an average of $0.5$ for the other tasks \citep{depasquale2018full}.
The parameter of the FORCE rule is $\alpha=1$. Dynamical learning
lasts for $t_{\text{learn}}$. During dynamical learning, we determine
$\bar{c}$ by averaging the context signal with an exponentially forgetting
kernel ($\tau_{\text{forget}}=50$ for task (v) and $\tau_{\text{forget}}=5$
for the other tasks). Testing lasts for $t_{\text{\text{test}}}$.

In all applications, recurrent weights $A_{ij}$ are set to zero with
probability $1-p$. Nonzero weights are drawn from a Gaussian distribution
with mean $0$ and variance $\frac{g^{2}}{pN}$, where $g=1.5$ \citep{SA09}.
We draw the feedback weights $w_{z,ij},$ $w_{c,ij}$ and the input
weights $w_{\varepsilon,ij},$ $w_{u,ij}$ from a uniform distribution
between $-\tilde{w}$ and $\tilde{w}$, set all initial output weights
$o_{z,ij}(0)$ and $o_{c,ij}(0)$ to $0$ and draw the biases $b_{i}$
from a uniform distribution between $-0.2$ and $0.2$. The number
of external inputs is $N_{u}.$ We use the standard Euler method for
our simulations, with an integration time step of $dt=0.1$, except
for Figs.~4 and \ref{fig:lorenzanalysis}, where we use $dt=0.01$
and $dt=0.025$, respectively. See \citep{Code19DynLearn} for example
code for task (i).

Further settings in the individual tasks are as follows:

Task (i): $N=500,N_{z}=1,N_{c}=1,N_{u}=0,p=0.1,\tilde{w}=1,t_{\text{stay}}=500,t_{\text{fb}}=100,t_{\text{wlearn}}=50000,t_{\text{learn}}=50,t_{\text{test}}=5000$.
The network learns to generate sinusoidal oscillations with period
$T$. The family of target trajectories is $\tilde{z}(t;T)=5\sin(\frac{2\pi}{T}t)$.
We use three different teacher trajectories for pretraining, with
periods $T=10,15,20$ and corresponding context targets $\tilde{c}=2,2.5,3$.
The target of dynamical learning in Figs.~2a and \ref{fig:learnspeed}
has $T=12.5$.

Task (ii): $N=500\text{\textendash}3000,N_{z}=1,N_{c}=1,N_{u}=0,p=0.1,\tilde{w}=1,t_{\text{stay}}=500,t_{\text{fb}}=100,t_{\text{wlearn}}=50000,t_{\text{learn}}=100,t_{\text{test}}=500$.
We do not update the output weights during a time interval of 20 at
the beginning of each training period. The network learns to generate
a superposition of two Fourier series with weighting factor $\lambda$.
The family of target trajectories is $\tilde{z}(t;\lambda)=(1-\lambda)\tilde{z}_{1}(t;\lambda)+\lambda\tilde{z}_{2}(t;\lambda)$
with $\tilde{z}_{l}(t)=\frac{1}{C_{l}}(\frac{\tilde{a}_{l,0}}{2}+\sum_{o=1}^{O}\tilde{a}_{l,o}\sin(\frac{2\pi o}{T(\lambda)}t+\tilde{\varphi}_{l,o}))$,
$l=1,2$, and $T(\lambda)=(1-\lambda)T_{1}+\lambda T_{2}$. We draw
the $\tilde{a}_{l,0}$, $\tilde{a}_{l,o}$, $\tilde{\varphi}_{l,o}$
and $T_{l}$ from uniform distributions between $-10$ and $10$,
$0$ and $10$, $0$ and $2\pi$, and $20$ and $50$, respectively.
$C_{l}$ is drawn from a uniform distribution to normalize the maximal
value of $|\tilde{z}_{l}(t)|$ to a random value between 3 and 7.
We use seven different teacher trajectories for pretraining, with
weighting factors $\lambda$ distributed equidistantly between 0 and
1. The corresponding context targets are distributed equidistantly
between 2 and 3. The target of dynamical learning in Fig.~2c has
$N=2000,O=6,\lambda=\frac{7}{12}$.

Task (iii): $N=1000,N_{z}=1,N_{c}=2,N_{u}=0,p=0.2,\tilde{w}=1,t_{\text{stay}}=500,t_{\text{fb}}=100,t_{\text{wlearn}}=50000,t_{\text{test}}=1000$.
The network learns to generate a superposition of sinusoidal oscillations
with amplitude $a$ and period $T$. The family of target trajectories
is $\tilde{z}(t;a,T)=a\left(\sin(\frac{2\pi}{T}t)+\cos(\frac{4\pi}{T}t)\right)$.
We use sixteen different teacher trajectories for pretraining, with
four amplitudes $a$ distributed equidistantly between 3 and 7 and
four periods $T$ distributed equidistantly between 10 and 20. The
corresponding context targets are distributed equidistantly between
2 and 3 for both parameters. The target of dynamical learning in Figs.~3a
and \ref{fig:learnspeed} has $a=5$ and $T=15$.

Task (iv): $N=500,N_{z}=3,N_{c}=1,N_{u}=0,p=0.1,\tilde{w}=1,t_{\text{stay}}=200,t_{\text{fb}}=100,t_{\text{wlearn}}=50000,t_{\text{learn}}=50,t_{\text{test}}=1000$.
The network learns to generate a constant output positioned on a curve
in three-dimensional space parameterized by $s$. The family of target
trajectories (fixed points) is $\tilde{z}(t;s)=\left(\frac{s^{3}}{2}+s_{\text{off}},2(s-\frac{1}{2})^{2}+s_{\text{off}},\frac{s}{2}+s_{\text{off}}\right)$,
where the offset $s_{\text{off}}=2.5$ ensures that the network feedback
is strong enough to entrain the reservoir network. We use ten different
teacher trajectories for pretraining with parameters $s$ chosen between
0 and 1 such that the corresponding $\tilde{z}(t;s)$ lie equidistantly
on the target curve $\left\{ \tilde{z}(t;s)|s\in[0,1]\right\} $.
The corresponding context targets are distributed equidistantly between
2 and 3. The targets of dynamical learning in Fig.~3b have $s=0.10$
and $s=0.92$.

Task (v): $N=1000,N_{z}=1,N_{c}=1,N_{u}=1,p=0.2,\tilde{w}=2,t_{\text{stay}}=1000,t_{\text{wlearn}}=30000,t_{\text{learn}}=200,t_{\text{test}}=500$.
We choose $\tau_{i}$ from a uniform distribution between $0.3$ and
$2.5$. During pretraining, we always provide error input $\varepsilon(t)$
to the network and do not fix $c(t)$, i.e. $t_{\text{fb}}=t_{\text{stay}}=1000$.
The network learns to predict the angle of a driven overdamped pendulum
with mass $m$. The family of target dynamical systems is given by
$\dot{\tilde{z}}(t)=F(\tilde{z}(t),u(t);m)=-m\sin(\tilde{z}(t))+u(t)-\exp((\tilde{z}(t)-0.65\pi)/0.65\pi)+\exp(-(\tilde{z}(t)+0.65\pi)/0.65\pi)$.
The last two terms provide a soft barrier preventing the pendulum
from undergoing full rotations. During pretraining and dynamical learning,
the pendulum is driven by low-pass filtered white noise $\dot{u}_{\text{wlearn}}(t)=-u_{\text{wlearn}}(t)+0.2dW/dt$
(see Fig.~\ref{fig:overdamped}b), which allows a comprehensive sampling
of the pendulum's dynamics. During testing the pendulum is driven
by a triangular wave with unit amplitude and period $T$= 50. We use
three different teacher dynamical systems for pretraining, with $m=0.5,1.0,1.5$
and corresponding context targets $\tilde{c}=0.7,0.95,1.2$. The targets
of dynamical learning in Fig.~3c,d have $m=0.8$ (continuous trace)
and $m=1.2$ (dashed trace).

Task (vi): $N=1000,N_{z}=3,N_{c}=1,N_{u}=0,p=0.1,\tilde{w}=2,t_{\text{stay}}=1000,t_{\text{fb}}=100,t_{\text{wlearn}}=50000,t_{\text{learn}}=50,t_{\text{test}}=10000$.
The network learns a Lorenz system with dissipation parameter $\beta$.
During pretraining, we always provide error input $\varepsilon(t)$
to the network, but fix $c(t)$ after $t_{\text{fb}}$. The family
of target dynamical systems is given by $\dot{\tilde{z}}(t)=F(\tilde{z}(t);\beta)=F_{\text{Lorenz}}(C_{\text{Lorenz}}\tilde{z}(t);\beta)/(C_{\text{Lorenz}}\tau_{\text{Lorenz}})$,
where $C_{\text{Lorenz}}=40$ and $\tau_{\text{Lorenz}}=20$ determine
the spatial and temporal scale of the dynamics and $F_{\text{Lorenz}}(x(t);\beta)=\left(\sigma(x_{2}-x_{1}),x_{1}(\rho-x_{3})-x_{2},x_{1}x_{2}-\beta x_{3}\right)$
is the vector field of the standard Lorenz system, with $\sigma=10$
and $\rho=70$. We use four teacher dynamical systems for pretraining,
with parameters $\beta$ distributed equidistantly between 2 and 6
and corresponding context targets distributed equidistantly between
2 and 3. The target of dynamical learning in Fig.~3e,f and \ref{fig:learnspeed}
has $\beta=4$.

\clearpage{}

\section{Quantification of learning performance\label{sec:Quantification-of-learning}}

To quantify the performance of our model, we measure for each application
the errors between signal outputs and targets during testing, for
different network instances and targets. Except for task (vi), we
compute the testing error as the root-mean-square error between signal
output and target during a period of length 50 in the middle of the
testing phase. The measure is chosen to ignore phase shifts that occur
over long testing times, as they are unavoidable in periodic autonomous
dynamics (tasks (i,ii)), due to the accumulation of small errors in
the period.

\begin{figure}[h]
\includegraphics{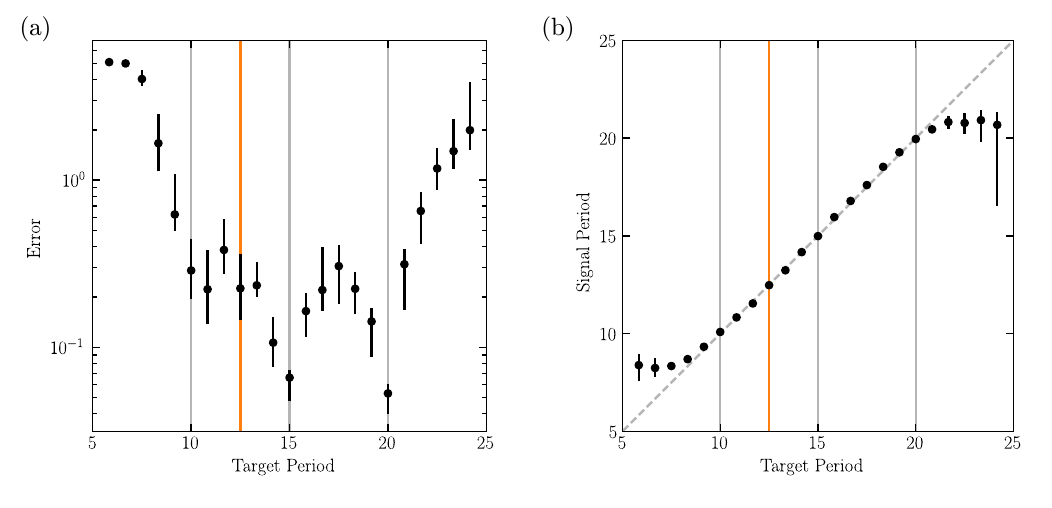}\caption{\label{fig:sine}Quality of dynamical learning of the sinusoidal oscillations
in task (i). (a) Testing error between signal output and target and
(b) period of the signal output, as a function of the period of the
target. Vertical gray lines indicate the periods of the pretrained
targets and vertical orange lines indicate the period of the target
used in Fig.~2a. Dots show median value and errorbars represent the
interquartile range between first and third quartile, using 10 network
instances.}
\end{figure}

Task (i): Fig.~\ref{fig:sine}a shows the testing error for the learning
of sinusoidal oscillations. It is small for targets with periods within
and slightly beyond the range spanned and interspersed by pretrained
targets. Fig.~\ref{fig:sine}b shows the good agreement between the
periods of the output signals and the targets. We determine the periods
from the maxima of the output signals' power spectra, after discarding
the initial interval of length $100$ of the testing phase to allow
for equilibration.

\begin{figure}[h]
\includegraphics{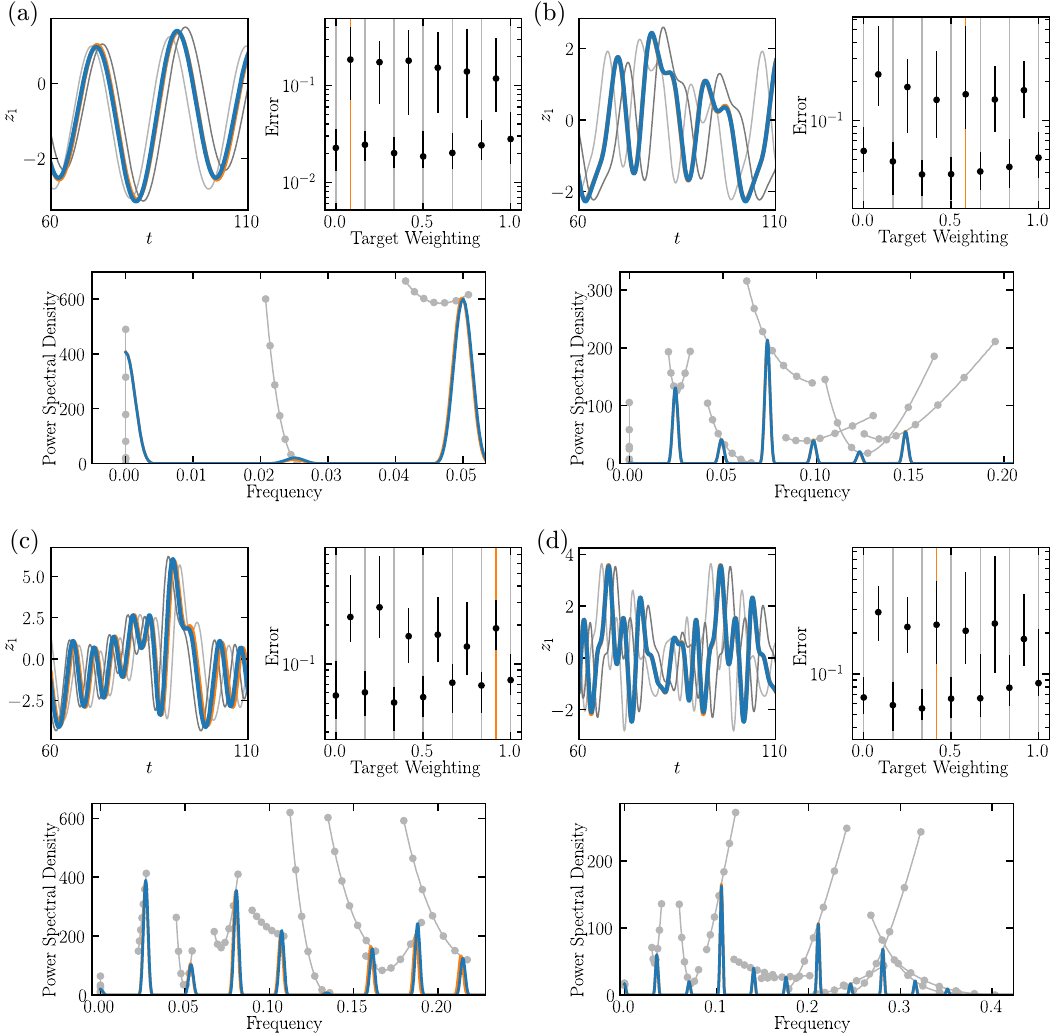}\caption{\label{fig:fourier}Quality of dynamical learning of the superposition
of Fourier series in task (ii). (a) $N=1000,O=2$ (a, top left) Signal
(blue) and target (orange) together with the two closest pretrained
dynamics (gray) during testing after dynamical learning of an unseen
target. (a, bottom) Power spectral density of signal (blue) and target
(orange) during testing. Gray lines show the power of the individual
Fourier components of the target family within the range of pretrained
targets. Gray dots indicate the targets used during pretraining. (a,
top right) Testing error between signal output and target as a function
of the target weighting factor. Vertical gray lines indicate the weighting
factors of the pretrained targets and vertical orange lines indicate
the weighting factors of the targets used in the other subpanels.
Dots show median value and errorbars represent the interquartile range
between first and third quartile, using 40 network instances and random
Fourier series. (b-d) Same as (a) but with (b) $N=2000,O=6$, (c)
$N=2500,O=8$, (d) $N=3000,O=10$.}
\end{figure}

Task (ii): Fig.~\ref{fig:fourier} shows the testing error for four
different combinations of network size $N$ and order $O$ together
with the signal and target in time and frequency domain for example
instances of task (ii), i.e., for specific realizations of the random
Fourier series described above. The testing error is low within the
range of pretrained weighting factors, especially for the targets
used during pretraining.

\begin{figure}[h]
\includegraphics{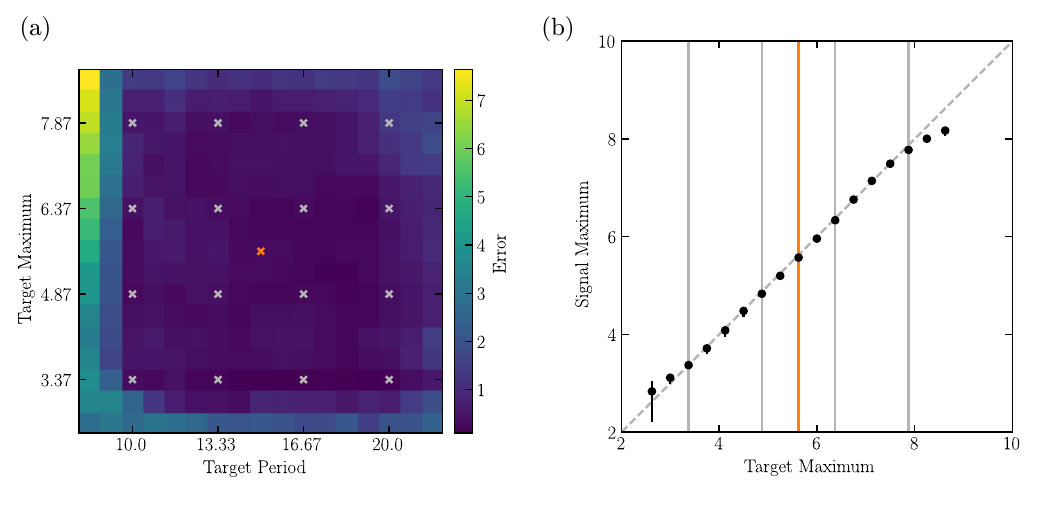}\caption{\label{fig:supersine}Quality of dynamical learning of the superpositions
of sines in task (iii). (a) Median testing error between signal output
and target as a function of the maximum and the period of the target
function. Gray crosses indicate parameters of the pretrained targets
and the orange cross indicates the parameters used in Fig.~3a. (b)
Averaged local maxima of the signal output as a function of the averaged
local maxima of the target, for a target period of $T=15$. Vertical
gray lines indicate the maxima of the pretrained targets and the vertical
orange line indicates the maximum of the target used in Fig.~3a.
Dots show median value and errorbars represent the interquartile range
between first and third quartile. Results in (a) and (b) are obtained
using 10 network instances for each parameter pair.}
\end{figure}

Task (iii): Fig.~\ref{fig:supersine}a shows the testing error for
the learning of superpositions of sines. Again, the error is low within
and slightly beyond the range of the parameters of the pretrained
targets. Similarly, the averaged local maxima of the signal outputs
agree well with the averaged local maxima of their targets, Fig.~\ref{fig:supersine}b.
The measurement of maxima starts at time 100 after the beginning of
testing.

\begin{figure}[h]
\includegraphics{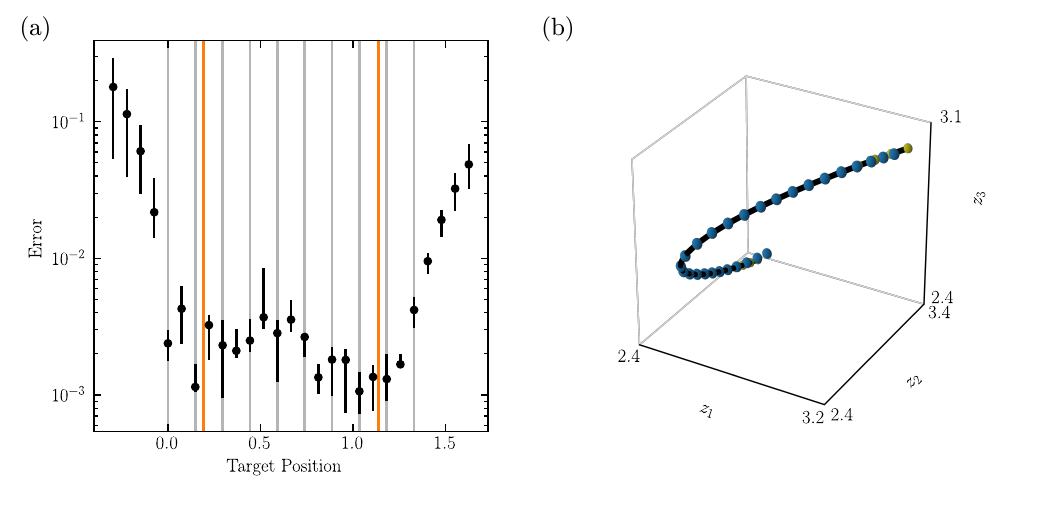}\caption{\label{fig:fixedpoint}Quality of dynamical learning of the fixed
points in task (iv). (a) Testing error between signal output and target
as a function of the target position. Vertical gray lines indicate
the positions of the pretrained targets and vertical orange lines
indicate the positions of the targets used in Fig.~3b. Dots show
median value and errorbars represent the interquartile range between
first and third quartile, using 10 network instances. (b) Single network
instance learning the same set of dynamical learning targets as in
(a). Blue spheres indicate the last signal outputs during testing
after the different instances of dynamical learning. Yellow spheres
indicate the position of the corresponding targets. They are mostly
covered by blue spheres, except in the regions of larger error. The
black tube shows the curve $\tilde{z}(t;s)$ on which the targets
lie.}
\end{figure}

Task (iv): Fig.~\ref{fig:fixedpoint}a shows the testing error for
the learning of fixed points. It is low for target positions within
and slightly beyond the range of the positions of the pretrained targets.
Fig.~\ref{fig:fixedpoint}b shows signal outputs for different targets
dynamically learned by a single network instance.

\begin{figure}[h]
\includegraphics{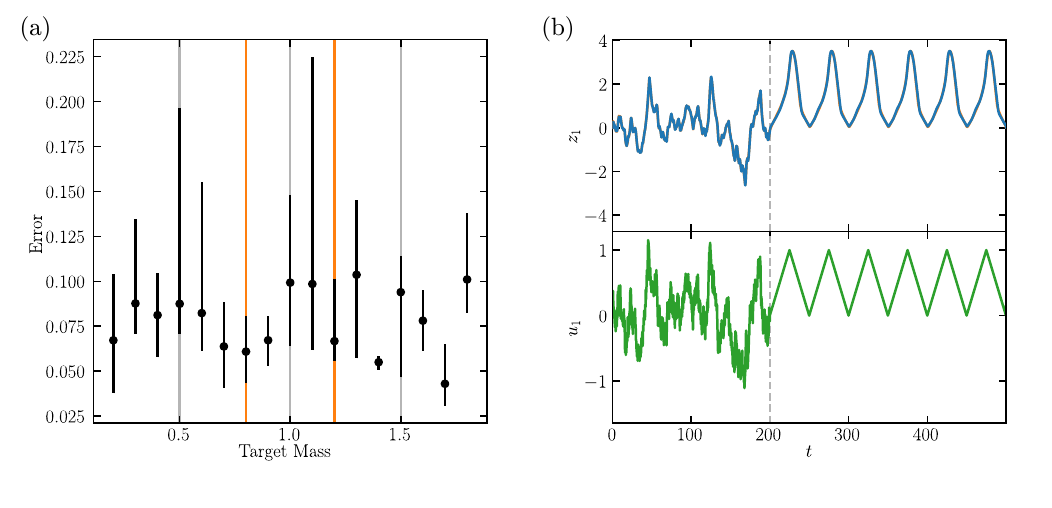}

\caption{\label{fig:overdamped}Quality of dynamical learning of the overdamped
pendulums in task (v). (a) Error between signal output and target,
as a function of the target pendulum's mass. Vertical gray lines indicate
the masses of the pretrained targets and vertical orange lines indicate
the masses of the targets used in Fig.~3d,e. Dots show median value
and errorbars represent the interquartile range between first and
third quartile, using 10 network instances. (b) Dynamical learning
and testing. The network and the target receive the same low-pass
filtered white noise as input drive during dynamical learning and
triangular wave input during testing (lower subpanel). The network
response (upper subpanel, blue trace) agrees well with the response
of the target (upper subpanel, orange trace, nearly completely covered
by the blue trace).}
\end{figure}

Task (v). Fig. \ref{fig:overdamped}a shows the testing error for
the learning of driven overdamped pendulums. It is small for pendulums
with masses within and slightly beyond the range spanned and interspersed
by pretrained pendulums. Fig.~\ref{fig:overdamped}b illustrates
the dynamical learning and testing phases.

\begin{figure}[h]
\includegraphics{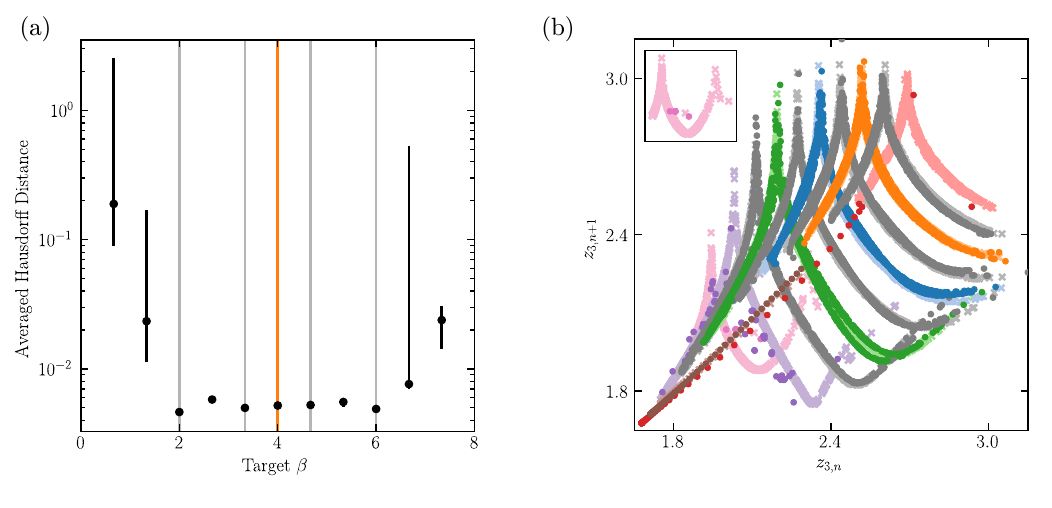}\caption{\label{fig:lorenz}Quality of dynamical learning of the Lorenz systems
in task (vi). (a) Testing error comparing the limit sets of signal
output and target, as a function of the target's parameter $\beta$.
Vertical gray lines indicate the parameters of the pretrained targets
and the vertical orange line indicates the parameter of the target
used in Fig.~3e,f. Dots show median value and errorbars represent
the interquartile range between first and third quartile, using 10
network instances. (b) Tent maps of subsequent maxima in the z-coordinate
for the signal output (dots, colored differently for different targets)
and for the target dynamics (crosses, light coloring alike corresponding
dots). The parameters $\beta$ of the targets are the same as in (a).
Dynamical learning of all targets with a single network instance.
Blue data correspond to the signal and target used in Fig.~3e,f;
gray data indicate pretrained targets. Tent maps of the target dynamics
move from bottom left to top right for increasing $\beta$ except
for the largest $\beta$ (brown, bottom left), for which the target
dynamics converge to a fixed point. Inset show close-up of results
for the smallest considered value of $\beta$. The signal output goes
to a fixed point for the two largest, but also for the smallest considered
value of $\beta$, leading to a focusing of the maxima relation to
a small region.}
\end{figure}

Task (vi): Since the Lorenz system is chaotic for most of the parameter
range that we consider, the signal output trajectory quickly deviates
from the target system's trajectory during testing. This holds also
if the network approximates the target dynamical system well. Hence,
instead of using the root-mean-square error, we compute the testing
error as the discrepancy of the limit set $M_{\text{net}}$ generated
by the network and the limit set $M_{\text{tar}}$ generated by the
target dynamics. For the comparison, we use the Averaged Hausdorff
Distance \citep{SchutzeELC12},
\begin{align}
d_{\text{AHD}}(M_{\text{net}},M_{\text{tar}})= & \max\left[\frac{1}{\mid M_{\text{net}}\mid}\sum_{m_{\text{net}}\in M_{\text{net}}}d(m_{net},M_{\text{tar}}),\frac{1}{\mid M_{\text{tar}}\mid}\sum_{m_{\text{tar}}\in M_{\text{tar}}}d(m_{tar},M_{\text{net}})\right],\\
d(m,M)= & \min_{m'\in M}\parallel m-m'\parallel,\nonumber 
\end{align}

which is robust against outliers. Fig.~\ref{fig:lorenz}a shows that
the testing error is low within the range of parameters $\beta$ spanned
and interspersed by pretrained targets. In addition, we find that
the relation between subsequent maxima of the z-coordinate of the
signal output correctly forms the shape of a tent for most tested
parameters (Fig.~\ref{fig:lorenz}b). The behavior of our model also
reproduces a bifurcation occurring for large $\beta$: The target
Lorenz system changes from chaotic behavior to fixed point behavior
for the largest value of $\beta$ we consider. Our networks dynamically
learn to generate the fixed point dynamics from this target, although
they were only pretrained in the chaotic regime. We note that some
network instances, for example the one shown in Fig.~\ref{fig:lorenz}b,
generate fixed point behavior during testing, if the target has the
second largest value of $\beta$ and is thus still chaotic. However,
also in these cases the signal output converges to one of the two
fixed points appearing for the largest $\beta$. This suggests that
due to a shift in the averaged context parameter, the dynamical regime
beyond the bifurcation is generated during testing.

\clearpage{}

\section{Analysis of dynamical learning of chaotic dynamics\label{sec:Analysis-lorenz}}

\begin{figure}[h]
\includegraphics{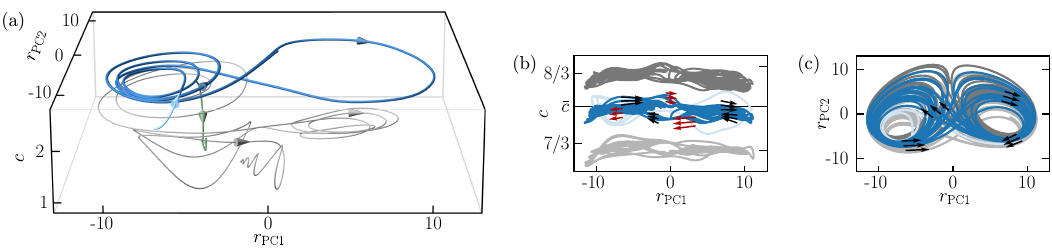}\caption{\label{fig:lorenzanalysis}Recurrent network dynamics during dynamical
learning (a) and testing (b,c) of task (vi), in $c,r_{\protect\PC1},r_{\protect\PC2}$-coordinates
(see Fig.~4 for task (i)). (a) During dynamical learning, the error
input drives the network to an orbit whose signal output approximates
the desired Lorenz system and keeps it there (light blue and blue
trajectories). Without input, the dynamics converge to a stable fixed
point, after a transient that yields a Lorenz system-like signal (gray).
Freezing $\tilde{z}(t)=\tilde{z}(t_{0})$ drives the dynamics quickly
to a fixed point (green). (b) During testing, the assumed orbit (blue)
resembles the error driven one in the $c\text{-}r_{\protect\PC,1}$-plane
(light blue, closest pretrained orbits with $c(t)$ fixed to their
$\tilde{c}$: gray). The constant feedback $\bar{c}$ prevents the
dynamics to leave the region where $c(t)\approx\bar{c}$, compare
$\dot{r}(r)$ (black vectors, $r$ on/nearby trajectory) with $\dot{r}(r)$
for variable feedback $c(t)$ (red vectors). (c) All four orbits are
similar in the $r_{\protect\PC,1}\text{-}r_{\protect\PC,2}$-plane,
since the dynamically learned orbit has a similar projected vector
field (black vectors) as the nearby pretrained ones.}
\end{figure}

To show that the mechanisms underlying dynamical learning and testing
that we worked out using task (i) in the main text also hold for a
qualitatively different, chaotic system, Fig.~\ref{fig:lorenzanalysis}
analyzes them for task (vi). As expected, we observe that during dynamical
learning, the error feedback drives the dynamics towards an orbit
generalizing the pretrained ones and keeps it there. During testing,
the network generalizes the pretrained characteristics to autoencode
$c$ such that the dynamics stay near the $\bar{c}$-plane in $r$-space
when the feedback $w_{c}c$ is clamped to $w_{c}\bar{c}$. The trajectory
is for task (vi) usually chaotic and generates the desired output
signal, because the vector field projected to the $c(t)=\bar{c}$-hyperplane
inter- or extrapolates nearby vector fields of other $c$-hyperplanes,
which embed pretrained orbits generating Lorenz dynamics with neighboring
parameters.

\clearpage{}

\section{Learning speed of dynamical learning\label{sec:Learning-speed}}

\begin{figure}[h]
\includegraphics{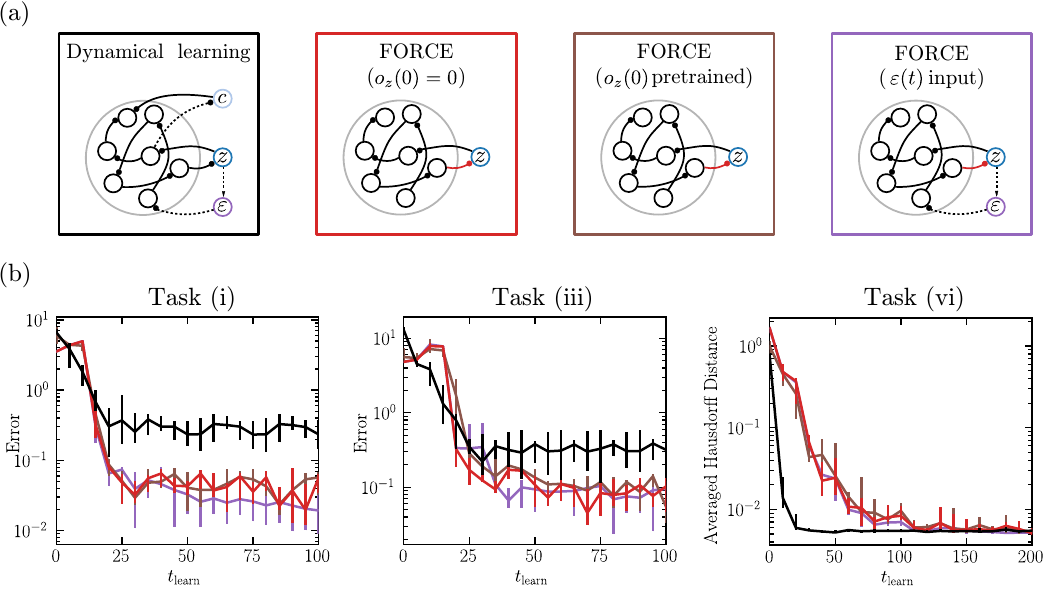}\caption{\label{fig:learnspeed}Learning speed of dynamical learning and FORCE
weight-learning. (a) Schematics of the different learning schemes.
Style of drawing has same meaning as in Fig.~1 of the main text.
(b) Testing error as a function of learning time for dynamical learning
(black), FORCE learning with $o_{z}$ initialized to zero (red), FORCE
learning with $o_{z}$ initialized to the signal output weights after
pretraining (brown) and FORCE learning with error input and $o_{z}$
initialized to 0 (purple, colors are alike frame colors in (a)). Connected
points represent median value and errorbars represent the interquartile
range between first and third quartile, using 10 network instances.}
\end{figure}

In the following we quantitatively assess the speed of dynamical learning.
We compare it with that of standard FORCE weight-learning, which uses
reservoirs with only a signal output $z(t)$ and output weight-learning.
As example tasks we consider learning of the sinusoidal oscillation,
main text task (i), of the superposition of sines, main text task
(iii), and of the Lorenz system, main text task (vi). The reservoirs
for standard FORCE learning have our standard parameters, except that
the biases are drawn from a uniform distribution between -5 and 5.
Further, the output weight-learning parameters are adapted; we apply
weight updates on every integration time step and set $\alpha$ to
$0.001$. Both changes improve performance and are for some combinations
of configuration and task even necessary for convergence. We consider
three different configurations of standard FORCE learning (Fig.~\ref{fig:learnspeed}a):
First, the typical configuration of a reservoir without input and
initialization of $o_{z}$ to 0. In the second configuration $o_{z}$
is initialized instead to the signal output weights obtained at the
end of pretraining for dynamical learning. This accounts for the possibility
that these output weights are beneficial initial conditions for weight-learning
and that our structural learning facilitates subsequent FORCE learning
despite the lack of context input, which was present during pretraining.
In the third configuration $o_{z}$ is initialized to $0$ and the
reservoir receives an error input $\varepsilon(t)=z(t)-\tilde{z}(t)$
during learning, because this might also facilitate FORCE learning.
To evaluate performance after different learning durations, we compute
testing errors as described in Sec.~\ref{sec:Quantification-of-learning}.
As usual, we stop weight modifications and, if present, error input
during testing. For a fair comparison, for dynamical learning with
$t_{\text{learn}}=0$ we fix the context to 0.

We find that dynamical learning is similarly fast or faster than FORCE
(Fig.~\ref{fig:learnspeed}b). For tasks (i) and (iii), both dynamical
learning and FORCE learning converge within approximately two periods
of the target dynamics ($T=12.5$ and $T=15$). FORCE learning converges
to smaller errors. For task (vi), dynamical learning converges in
about five cycles (maxima of the $z$-coordinate) of the target system.
FORCE learning is about five times slower and yields similar errors.
The similar convergence speed of the first two configurations in all
considered tasks indicates that FORCE weight-learning does not profit
from our form of pretraining.

Taken together, we observe that dynamical learning converges within
a few characteristic timescales of the target dynamics and is thus
on par with FORCE learning for simple and faster converging for complex
tasks. This held for both the standard and the hand-tuned parameter
sets. The observation is plausible since for complicated tasks FORCE
learning needs to gather information that dynamical learning already
possesses due to the previous pretraining. It is especially interesting
because dynamical learning may be considered biologically plausible
and because FORCE is a recommended reservoir computing scheme \citep{lukosevicius2012reservoir}.

\clearpage{}

\section{Robustness of learning performance\label{sec:Range-target}}

To check the robustness of our dynamical learning scheme against changes
in task family parameters, we determine its performance for different
families of sinusoidal oscillations, main text task (i). Specifically,
we vary the number of pretrained targets, the amplitude of the oscillations,
the difference between the maximal and minimal period of the pretrained
targets (period range) as well as the minimal period of the weight-learned
targets. For each combination of these task family parameters, we
pretrain the networks as before. Afterwards, we dynamically learn
a set of targets with periods ranging from the smallest to the largest
pretrained period, where the period increases by one between neighboring
targets. We compute a normalized error for each target and take the
average to quantify the performance of the network for the considered
task family. The normalized error is the root-mean-square error during
a period in the middle of the testing phase, with length three times
the target period, divided by the corresponding root-mean-square error
assuming that the signal output is zero.

\begin{figure}[h]
\includegraphics{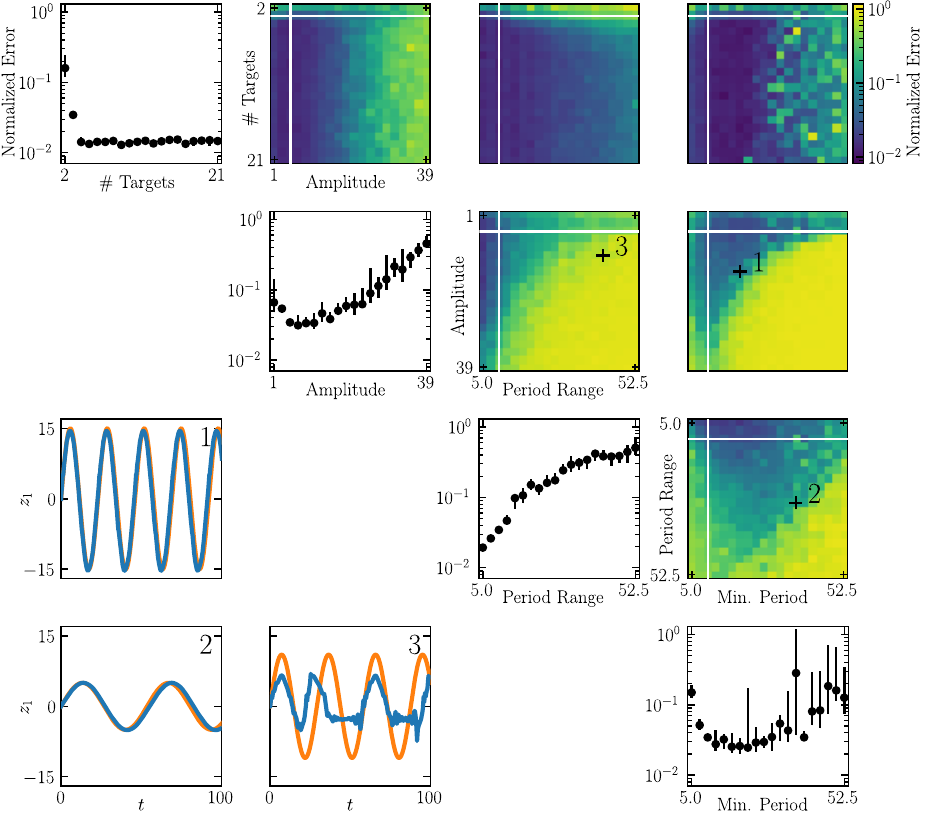}\caption{\label{fig:quantperf500}Performance over a broad range of task family
parameters. Panels on and above the diagonal show the average normalized
errors taken over sets of testing targets. All but the indicated parameters
are set to their standard values. White lines in panels above the
diagonal indicate the parameter values of the one-dimensional slices
shown on the diagonal. Dots and color represent median value and errorbars
in panels on the diagonal represent the interquartile range between
first and third quartile, using 10 network instances. Panels below
the diagonal show representative dynamically learned example signal
outputs (blue) and corresponding targets (orange) for the three different
parameter combinations indicated by numbered crosses in the panels
above the diagonal.}
\end{figure}

\begin{figure}[h]
\includegraphics{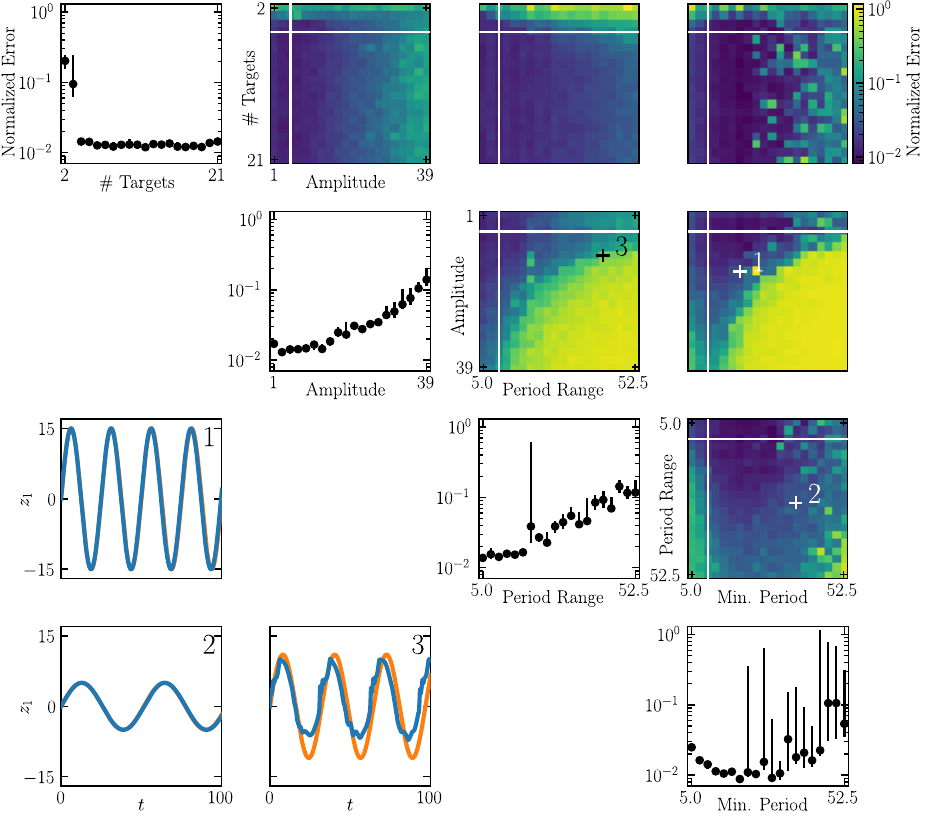}\caption{\label{fig:quantperf1000}Same as Fig.~\ref{fig:quantperf500} for
networks with 1000 neurons and five pretrained targets unless the
number of pretrained targets is varied.}
\end{figure}

To compute and interpret the errors in high-dimensional parameter
space, we cut out slices where we keep all but at most two of the
task family parameters at their standard values specified in Sec.~\ref{sec:Additional-detail}.
We find that dynamical learning works robustly for large parameter
regions. In particular, the number of targets and the period range
can often be changed over an order of magnitude, see Fig.~\ref{fig:quantperf500}.
Increasing the network size to 1000 neurons and the number of pretrained
targets to five instead of three further increases robustness against
changing other parameters, see Fig.~\ref{fig:quantperf1000}. Taken
together, we may conclude that our scheme works well for a wide range
of task families.

\clearpage{}

\section{Induction of unseen signal outputs by a context-like external input\label{sec:Induction-of-unseen}}

\begin{figure}[h]
\includegraphics{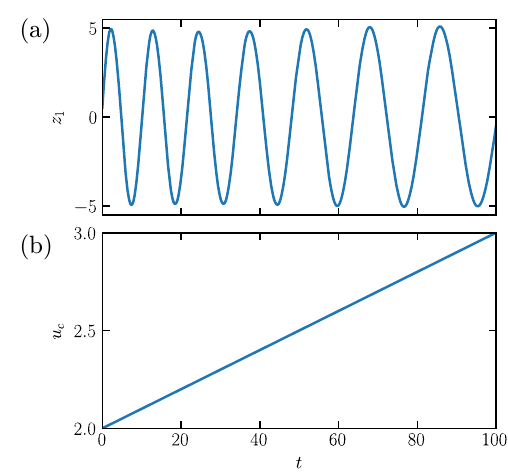}\caption{\label{fig:iogeneral}Induction of unseen signal outputs by a context-like
external input. The network has been trained similar to pretraining
in task (i) to generate sinusoidal oscillations with three different
frequencies in response to three constant external context inputs
$u_{c}(t)$. After training, the weights are fixed and the network
receives a continuously rising $u_{c}(t)$ (b). This results in a
sinusoidal signal output with continuously rising period, which interpolates
between the trained signals (a).}
\end{figure}

We test whether changing a context-like input $u_{c}(t)$ allows to
generate sinusoidal oscillations with previously unseen frequencies.
Like $c(t)$, $u_{c}(t)$ connects to the neurons in the network with
a weight matrix $w_{c}$. However, $u_{c}(t)$ is never generated
by a network output, but a purely external input. There is no further
context variable $c(t)$ and no error input $\varepsilon(t)$ in the
network. Apart from this, the network is setup like in task (i). The
output weights $w_{z}$ are learned using the FORCE rule, similar
to pretraining in task (i): during each training period, we teach
the network to generate a sinusoidal oscillation $\tilde{z}(t;T)$
with a period $T=10,15,20$, in response to a constant $u_{c}(t)=2,2.5,3$,
analogous to teacher forcing with $\tilde{c}$. We find that the system
can interpolate between the pretrained output signals, if driven by
previously unseen $u_{c}(t)$, cf.~Fig.~\ref{fig:iogeneral}. See
ref.~\citep{jaegerControllingRecurrentNeural2014} for a similar
finding when morphing between conceptor weight matrices. (The recent
`conceptor' approach fixes reservoir dynamics by weight changes \citet{jaegerControllingRecurrentNeural2014,jaegerUsingConceptorsManage2017}.)

\clearpage{}

\section{Pretraining with weight perturbation\label{sec:Pretraining-with-weigh}}

\textit{Introduction.} Throughout the article reservoir computing
with FORCE learning is used for pretraining, see main text and Supplemental
Material Secs.~\ref{sec:Reservoir-computing-and}-\ref{sec:Induction-of-unseen}.
In the following we show dynamical learning of simple tasks in networks
that are pretrained with a biologically more plausible rule. Specifically,
we use reservoir computing with weight perturbation \citep{Dembo1990,Jabri1992,Cauwenberghs1993,Hs2003}
to learn network structures that enable the dynamical learning of
fixed points in two-dimensional space. We note that the direct application
of a recent node perturbation scheme \citep{Miconi2017} to the output
or all neurons was hindered by difficulties with learning multiple
targets (cf.~also \citep{beer2018step}). Weight perturbation is,
in short, a local reinforcement learning rule that consists of three
steps: (i) randomly perturbing the connection weights, (ii) comparing
the obtained reward with the reward expected without perturbation,
and (iii) changing the connection weights into the direction (opposite
direction) of the perturbation if the actual reward is higher (lower)
than the expected one.

\textit{Structure learning with weight perturbation.} We use batch
learning, i.e the pretraining phase consists of $N_{\text{trials}}$
trials, each of which is comprised of the presentation of all $N_{\text{tar}}$
pretraining members of the task family $\tilde{z}(t;s)$ for a time
$t_{\text{stay}}$. The signal output weights $o_{z}$ learn as follows:
At the beginning of trial $n$, the weights $o_{z,ki}(n-1)$ from
the end of the previous trial receive small perturbations $\Delta o_{z,ki}^{\text{pert}}(n)$
\citep{Cauwenberghs1993}. The perturbations are drawn from a normal
distribution with zero mean and standard deviation $\sigma$. We define
the reward $R_{z}(n)$ as the negative sum of the mean squared errors
between the signals and their targets during an evaluation period
that starts $t_{\text{off}}$ after the beginning of the trial. Further,
we approximate the reward of the unperturbed network on the training
batch by an exponentially weighted average $\bar{R}_{z}(n-1)=\alpha\bar{R}_{z}(n-2)+(1-\alpha)R_{z}(n-1)$
of previous rewards with timescale $\alpha$ \citep{Miconi2017}.
This gives the estimate
\begin{equation}
G_{ki}(n)=\frac{\Delta o_{z,ki}^{\text{pert}}(n)}{\sigma^{2}}(R_{z}(n)-\bar{R}_{z}(n-1))
\end{equation}
for the weight gradient \citep{Cauwenberghs1993}. When we obtain
the weight updates $\Delta o_{z,ki}(n)$ directly from this estimate,
in our model we observe poor performance. It improves markedly when
we combine the estimate with the Adam algorithm \citep{Kingma2014}.
Adam introduces a momentum term $v_{ki}(n)$ and an individual learning
rate $1/\sqrt{g_{ki}(n)+\mu}$ for each connection such that our weight
update equations read
\begin{align}
o_{z,ki}(n)= & o_{z,ki}(n-1)+\Delta o_{z,ki}(n),\\
\Delta o_{z,ki}(n)= & \eta\frac{v_{ki}(n)}{\sqrt{g_{ki}(n)+\mu}},\\
v_{ki}(n)= & \beta v_{ki}(n-1)+(1-\beta)G_{ki}(n),\\
g_{ki}(n)= & \gamma g_{ki}(n-1)+(1-\gamma)G_{ki}^{2}(n).
\end{align}
Here, $\eta$ is the global learning rate, $\mu$ a constant preventing
overly large weight updates and $\beta$ and $\gamma$ are the timescales
of the exponential averaging of the momenta and the learning rates,
respectively. Learning of the context output weights is implemented
likewise.

\begin{figure}[h]
\includegraphics{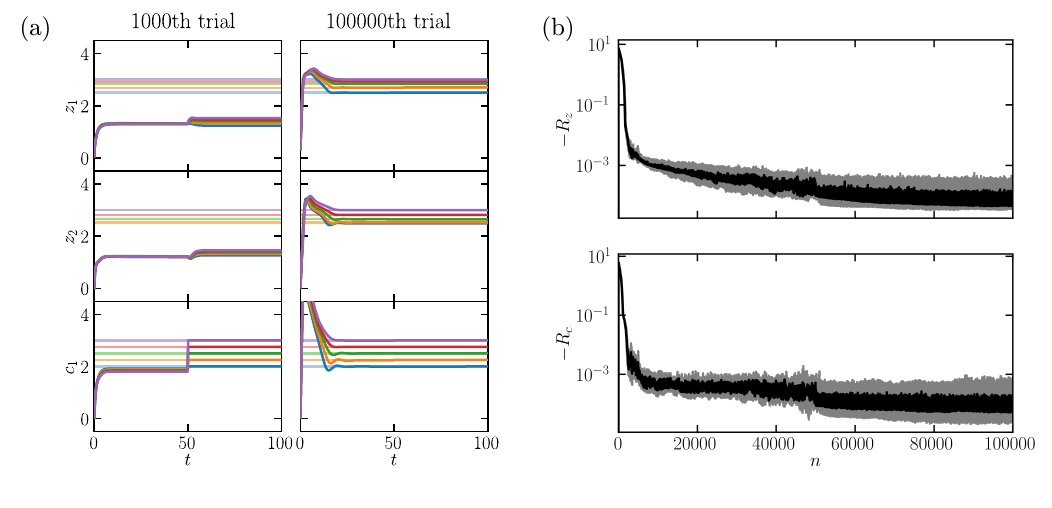}\caption{\label{fig:wplearn}Pretraining using weight perturbation. (a) Signal
and context outputs for the different batch members early (left) and
late (right) during pretraining for a single network instance. In
late trials the networks' error input induces a quick convergence
of the outputs (strong colors) to their targets (light colors). At
$t=50$, the context variable is fixed to its desired value. (b) Negative
reward for the signal (top) and context (bottom) output. Black line
shows median value of 10 network instances and gray area indicates
interquartile range between first and third quartile.}
\end{figure}

\begin{figure}[h]
\includegraphics{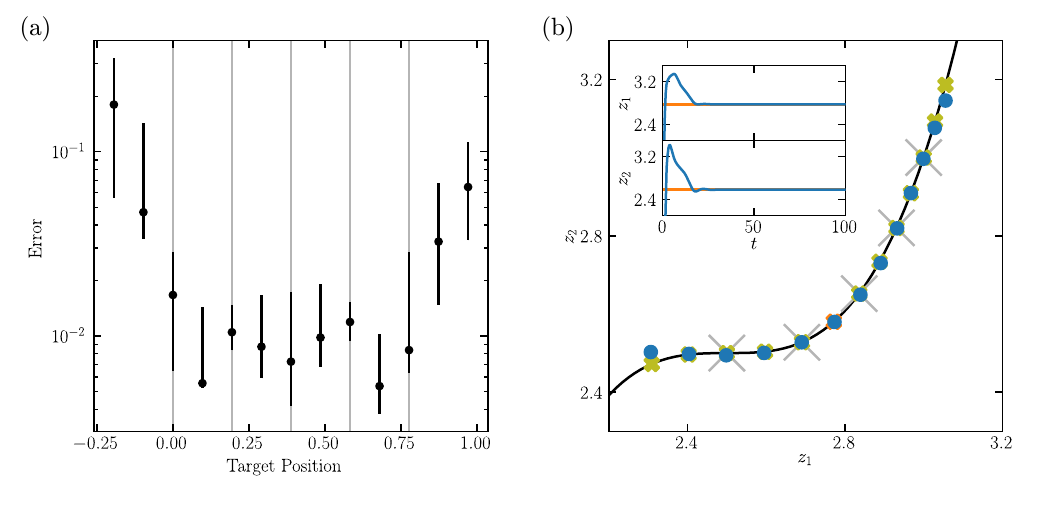}\caption{\label{fig:wptest}Quality of dynamical learning after pretraining
with weight perturbation. (a) Testing error between signal output
and target as a function of the target position. Vertical gray lines
indicate the positions of the pretrained targets. Dots show median
value and errorbars represent the interquartile range between first
and third quartile, using 10 network instances. (b) Single network
instance learning the same set of targets as in (a). The inset shows
signal outputs (blue) versus time during dynamical learning and truncated
testing periods for one of the targets (orange). In the main panel
blue dots indicate the last signal outputs during testing. Yellow
crosses indicate the position of the corresponding targets. They are
mostly covered by blue dots, except in the regions of larger error.
The orange cross indicates the position of the target shown in the
inset. Gray crosses indicate the positions of the pretrained targets.
The black line shows the curve $\tilde{z}(t;s)$ on which the targets
lie.}
\end{figure}

\textit{Results.} At the end of the pretraining trials, our networks
have learned to produce (in response to the signal error input) signal
and context outputs that are close to the pretrained targets within
the evaluation period, see Fig.~\ref{fig:wplearn}. The established
underlying network structures also enable the network to dynamically
learn previously unseen targets. Like for the differently pretrained
networks in the main text and Supplemental Material Secs.~\ref{sec:Reservoir-computing-and}-\ref{sec:Induction-of-unseen},
a short presentation of the (here constant) target signal via the
error input teaches the network to imitate it and to choose an appropriate
context. During a subsequent testing period, the network autonomously
continues the desired signal stabilized by the fixated context. Fig.~\ref{fig:wptest}a
shows the testing error after dynamical learning of different signals.
It is low for target positions within and slightly beyond the range
of the pretrained targets. Fig.~\ref{fig:wptest}b shows signal outputs,
which were dynamically learned by a single network instance.

\textit{Discussion. }We have shown that for a simple task pretraining
can also be performed with a learning rule that satisfies main criteria
for biological plausibility, as it is local and causal. Further, it
relies on delayed, sparse rewards and updates the weights at a low
rate at the end of a trial. It is biologically plausible that synapses
tentatively change their weights and then consolidate or reverse the
change, depending on reward \citep{redondo2010making}. To improve
learning, we have employed momentum and individual, history dependent
learning rates for each connection. Supported by experimental findings
it has already been argued that the brain could realize learning with
momentum \citep{Yu2016}. Furthermore there is ample evidence for
a complex history dependence of learning rates in individual synapses
\citep{abraham2008metaplasticity}. While our weight modifications
do not rely on a continuous supervisory signal anymore, such a signal
is still present in the error input, like during dynamical learning.
Future work may investigate how it can be replaced by sparse supervision.

\textit{Task details. }$N=1000,N_{z}=2,N_{c}=1,N_{u}=0,p=0.1,\tilde{w}=1,t_{\text{stay}}=100,t_{\text{fb}}=50,t_{\text{learn}}=50,t_{\text{test}}=1000,t_{\text{off}}=25,N_{\text{trials}}=10^{5},N_{\text{tar}}=5,\sigma=10^{-4},\alpha=1/3,\beta=0.99,\gamma=0.99,\mu=10^{-8}$,
$\eta=50\times10^{-5}$ for the first 5000 trials, $\eta=10\times10^{-5}$
for trials 5000 to 50000, $\eta=1\times10^{-5}$ afterwards. The model
and application details described in the main text and in Sec.~\ref{sec:Additional-detail}
also apply to the current setting, except for those concerning the
weight-learning rule. At the beginning of each pretraining trial for
each member of the batch we draw the initial activation variables
$x_{i}$ from a uniform distribution between $-0.1$ and $0.1$. The
network learns to generate a constant output positioned on a curve
in two-dimensional space parameterized by $s$. The family of target
trajectories (fixed points) is $\tilde{z}(t;s)=\left(\frac{s^{3}}{2}+s_{\text{off}},\frac{s}{2}+s_{\text{off}}\right)$,
where the offset $s_{\text{off}}=2.5$ ensures that the network feedback
is strong enough to entrain the reservoir network. We use four different
teacher trajectories for pretraining with parameters $s$ chosen between
0 and 1 such that the corresponding $\tilde{z}(t;s)$ lie equidistantly
on the target curve $\left\{ \tilde{z}(t;s)|s\in[0,1]\right\} $.
The corresponding context targets are distributed equidistantly between
2 and 3.


\clearpage{}

\section{Supplementary discussion\label{sec:Supp-Disc}}

We conclude with an extended discussion of our findings, previous
literature and possible future applications. Overall, we have shown
how neural networks can quickly learn trajectories and dynamical systems
without changing their weights and without requiring a teacher during
testing. During the pretraining (learning-to-learn), the networks
are taught several dynamics from the same family as the later dynamically
learned ones, as well as a corresponding constant context. The process
is supervised by an error signal to the synapses and, part of the
time, by an error input to the network. During dynamical learning,
a short presentation of the latter alone suffices to teach the desired
dynamics. The network then also generates a context, which fluctuates
around some temporal mean. When subsequently testing the generation
of the dynamics, the error input is removed and the context is fixed
to its average, telling that the learned dynamics should be continued.

Our analysis indicates that the scheme works due to an interplay of
generalization and stabilization: During pretraining, the networks
adapt to perform a negative feedback/autoencoder task. During dynamical
learning, they generalize this, by generating a new desired signal
when receiving its error as input. Simultaneously they choose a consistent
context. During testing, this context is externally kept constant,
which stabilizes the learned signal. This is possible because a mutual
association between contexts and targets has been weight-learned during
pretraining.

Approaches to supervised dynamical learning in the literature consider
the one-step prediction of time series \citep{feldkampAdaptationFixedWeight1996,feldkampAdaptiveBehaviorFixed1997}
and input-output maps \citep{youngerFixedweightOnlineLearning1999,hochreiterLearningLearnUsing2001a,youngerMetalearningBackpropagation2001,santiagoContextDiscerningMultifunction2004,santoroMetaLearningMemoryAugmentedNeural2016,bellecLongShorttermMemory2018},
where the correct previous output is fed in. Other networks could
adapt to provide negative feedback for control \citep{puskoriusNeurocontrolNonlinearDynamical1994,feldkampTrainingControllersRobustness1994,feldkampFixedweightControllerMultiple1997,oubbatiNeuralFrameworkAdaptive2010},
a pretrained oscillation \citep{wyffels2014frequency}, periodic sequences
of discrete states \citep{jaeger2008cant} or the parameters of a
dynamical system \citep{oubbatiMetaLearningAdaptiveIdentification2005}.
The studies use simple recurrent neural networks \citep{feldkampTrainingControllersRobustness1994,puskoriusNeurocontrolNonlinearDynamical1994,feldkampAdaptationFixedWeight1996,feldkampAdaptiveBehaviorFixed1997,feldkampFixedweightControllerMultiple1997,santiagoContextDiscerningMultifunction2004,oubbatiMetaLearningAdaptiveIdentification2005,oubbatiNeuralFrameworkAdaptive2010},
gated \citep{youngerMetalearningBackpropagation2001,hochreiterLearningLearnUsing2001a,santoroMetaLearningMemoryAugmentedNeural2016}
or spiking ones \citep{bellecLongShorttermMemory2018}, trained by
backpropagation \citep{hochreiterLearningLearnUsing2001a,youngerMetalearningBackpropagation2001,santoroMetaLearningMemoryAugmentedNeural2016,bellecLongShorttermMemory2018}
or extended Kalman filtering \citep{feldkampTrainingRobustNeural1994,puskoriusNeurocontrolNonlinearDynamical1994,feldkampTrainingControllersRobustness1994,feldkampAdaptationFixedWeight1996,feldkampAdaptiveBehaviorFixed1997,feldkampFixedweightControllerMultiple1997,feldkampSimpleConditionedAdaptive2003,oubbatiMetaLearningAdaptiveIdentification2005,oubbatiNeuralFrameworkAdaptive2010}.
The simple networks are similar to ours but use non-leaky neurons,
different learning and often assume discrete time. To our knowledge,
all the systems with continuous signal space were fed a form of the
temporally variable teaching signal also during testing.

Earlier work showed that sufficiently large recurrent neural networks
with static weights can approximate any smooth input-output dynamics
relation with bound-restricted inputs for finite time \citep{sontagNeuralNetsSystems1992,funahashi1993,kimStandardRepresentationUnified2018}.
This implies that a network with static weights can in principle approximate
the output of another, weight-learning one. The static network's dynamics
thereby include the effects of the other network's learning algorithm
and thus learns dynamically \citep{cotterFixedweightNetworksCan1990,cotterLearningAlgorithmsFixed1991}.
Our networks with static weights are not pretrained to approximate
during dynamical learning the outputs of weight-learning networks.
In particular they do not approximate the outputs of a FORCE weight-learning
reservoir computer, as illustrated by the different convergence properties
in Suppl.-Fig.~\ref{fig:learnspeed}.

In our networks, fixing the intrinsically chosen context $c(t)$ indicates
that the dynamics are to be continued. This is analogous to fixing
the weights during testing in weight-learning paradigms. It is necessary
to avoid convergence to other dynamics (if the system has discrete
attractors) or diffusion (for marginally stable dynamics). The instruction
to fix $c(t)$ is independent of the task and much simpler than task
specific teacher and target signals. The constant $c(t)$ can be stored
and kept up by biologically plausible circuits \citep{lim2013balanced}.
For long times, weight-learning may consolidate it. $c(t)$ may be
understood as (continuous) memory variable \citep{maass2007computational,sussillo2013opening}.
In contrast to previous ones it is neither a pure feedback output
\citep{maass2007computational,pascanu2011neurodynamical} nor an external
input (cf.~Sec\@.~\ref{sec:Induction-of-unseen} and \citep{mante2013context})
and it does not facilitate weight-learning \citep{maass2007computational,pascanu2011neurodynamical,mante2013context}.
One can also drive networks with external input such that unseen,
interpolating input leads to interpolating dynamics (cf.~Sec\@.~\ref{sec:Induction-of-unseen}
and \citep{bostrom2013model}). In contrast to such generalization,
our networks learn their new dynamics by imitating a teacher. In particular,
they adopt the phase of an oscillatory target.

Our pretraining implements a form of structure learning \citep{lansdellLearningtolearn2018,Braun2010},
i.e.~learning of the structure (concepts) underlying a family of
tasks, which in general facilitates subsequent learning of new representatives.
In our networks it enables learning of representatives without synaptic
modification. Experiments indicate that animals and humans employ
structure learning for example for motor tasks, which requires presentation
of a variety of representative tasks and involves a reduction of the
dimensionality of the search space, as in our model \citep{lansdellLearningtolearn2018,Braun2010}.
Evolution or network plasticity should implement structure learning
in biology. Since their functioning is largely unknown, we employ
a simple reservoir computing scheme and comparably small neural networks.
Only the readout weights, a small fraction of the network weights,
are trained. Our dynamically learned tasks have similar difficulty
as those used to introduce FORCE weight learning \citep{SA09}. They
are low-dimensional; this may often be the relevant case for biological
neural networks, e.g., when learning movements \citep{gallego2017neural}.

In experimental physics and engineering, our scheme may find application
in neuromorphic computing. Here, intrinsically plastic weights are
costly and often difficult to realize, while outsourcing the learning
to external controllers introduces computational bottlenecks \citep{chicca2014neuromorphic}.
As an example, in analog, photonic neuromorphic computing, network
weights are externally set to generate desired output dynamics \citep{duport2016fully,tait2017neuromorphic,antonik2017braininspired}.
Our scheme may allow such systems to intrinsically learn and thereby
fully reap their speed benefits. For spiking hardware, our networks
may be efficiently translated into spiking ones \citep{thalmeier2015universal}.
Dynamical learning may reduce the size and power consumption of such
hardware, for example in autonomous robots \citep{schuman2017survey}.

Our approach suggests a new method for the prediction of chaotic systems
\citep{pathak2018,zimmermann2018}, which searches for similarity
within a predefined family of dynamics and leaves the networks structurally
invariant and flexible.

A possible example for dynamical learning in biology is the quick
learning of new movements \citep{Braun2010,perich2018neural}, perhaps
with subsequent consolidation by plasticity. Another example may be
short-term memory of single items and temporal sequences \citep{sreenivasan2019delay,oberauer2019rehearsal}.
Our theory predicts that even complicated dynamics may be memorized
in biological neural networks without synaptic modification.

%